\documentstyle[aps,preprint,epsf,tighten]{revtex}

\def\p{{\bf p}}
\def\q{{\bf q}}
\def\k{{\bf k}}
\def\md{m_{\text{D}}}

\begin{document}
\setcounter{page}{0}
\def\footnoterule{\kern-3pt \hrule width\hsize \kern3pt}

\title{Superconductivity by long-range color magnetic interaction in
high-density quark matter\thanks{This work is supported in part by funds
provided by the U.S.  Department of Energy (D.O.E.) under cooperative
research agreement \#DF-FC02-94ER40818.}}

\author{D.T.~Son\footnote{Email address: {\tt son@ctp.mit.edu}}}
\address{Center for Theoretical Physics,
Laboratory for Nuclear Science
and Department of Physics \\
Massachusetts Institute of Technology,
Cambridge, Massachusetts 02139 \\
{~}}
\date{MIT-CTP-2805,~ hep-ph/9812287. December 1998}

\maketitle

\thispagestyle{empty}

\begin{abstract}

We argue that in quark matter at high densities, the color magnetic field
remains unscreened and leads to the phenomenon of color superconductivity. 
Using the renormalization group near the Fermi surface, we find that the
long-range nature of the magnetic interaction changes the asymptotic
behavior of the gap $\Delta$ at large chemical potential $\mu$
qualitatively.  We find $\Delta\sim\mu
g^{-5}\exp(-{3\pi^2\over\sqrt{2}}\cdot{1\over g})$, where $g$ is the small
gauge coupling.  We discuss the possibility of breaking rotational
symmetry by the formation of a condensate with nonzero angular momentum,
as well as interesting parallels to some condensed matter systems with
long-range forces. 

\end{abstract}
\vspace*{\fill}
\begin{center}
Submitted to: {\it Physical Review D}
\end{center}

\section{Introduction}

The study of QCD at finite density has a long history.  The suggestion
that at high densities hadronic matter becomes quark matter \cite{cp} was
made almost immediately after the discovery of asymptotic freedom
\cite{gwp}.  It has been known for almost as long that at very high baryon
densities, where perturbative QCD can be applied, quark matter is in a
BCS-type superconducting state \cite{BailinLove}.  More recently, various
groups \cite{ARW,RSSV} have modeled quark matter at intermediate
densities using phenomenological four-fermion interactions and found that
condensates form with sizable gaps of order 100 MeV.  These studies reveal
a rich phenomenology \cite{ARW,RSSV,Berges,locking,SchaeferWilczek},
including an interesting phase diagram \cite{Berges}, new symmetry
breaking schemes like color-flavor locking \cite{locking}, which might be
also relevant for the study of nuclear matter in the nuclear superfluid
phase, if the latter is continuously connected to the quark matter phase
\cite{SchaeferWilczek}. 

Naturally, a systematic calculation in perturbation theory is possible
only in the high-density regime, where the chemical potential $\mu$ is
much larger than $\Lambda_{\text{QCD}}$ and the strong coupling at the
scale of the Fermi energy is small.  However, even in this case reliable
calculations have so far been hampered by the inability to take into
account the long-range nature of the color magnetic force, which dominates
the interaction between the quarks at large distances.  Instead of
tackling the problem of condensate formation by a long-range interaction,
many treatments rely on the assumption that a magnetic mass of order
$g^2\mu$ is somehow generated.  However, in reality, the magnetic field
remains unscreened in the absence of superconductivity itself. Ordinary
BCS theory cannot be directly applied due to the IR divergence from the
exchange of the soft magnetic gluons.  Therefore, even the asymptotic,
weak-coupling behavior of the gap has not been found. 

In this paper we present what we believe to be the correct estimate for
the value of the BCS gap at asymptotically high densities.  Our approach
is based on the renormalization group around the Fermi surface
\cite{RG,EHS}, properly modified to take into account the long-range
magnetic interaction.  We find that the gap is proportional to
$\exp(-c/g)$, $c={3\pi^2\over\sqrt{2}}$.\footnote{This agrees with a
comment in a recent paper by Pisarski and Rischke \cite{Rischke} that
$\Delta\sim\exp(-c/g)$ for some constant $c$.} This behavior is different
from the naive expectation from BCS theory, which predicts $\exp(-c/g^2)$.
The fact that the suppression is parametrically much milder means a larger
value of the gap at high densities, and potentially could also indicate
that at intermediate densities the gap and the critical temperature may be
larger than previously estimated.  The latter may substantially enhance
the chance of forming a color superconductor in heavy-ion collisions.  We
also show that the asymptotic behavior of the gap does not depend on the
angular momentum of the Cooper pairs, and comment on the possibility of
breaking rotational symmetry by the formation of a condensate with a
nonzero angular momentum.

We will first review the renormalization group approach to the BCS theory
(Sec.\ \ref{sec:RG}), then describe the trouble caused by the long-range
magnetic interaction (Sec.\ \ref{sec:mag}).  In Sec.\ \ref{sec:core} we
describe our resolution of this problem.  We then make final remarks in
Sec.\ \ref{sec:conclusion}.  The Appendices contain various technical
details, including a treatment of Eliashberg theory.

\section{The renormalization group near the Fermi surface}
\label{sec:RG}

An elegant method to see the formation of the BCS superconducting state is
the renormalization group (RG) near the Fermi surface \cite{RG}.  This
approach has been applied to the case of quark matter by Evans, Hsu and
Schwetz, and Sch\"afer and Wilczek \cite{EHS}.  These treatments apply
when there exists a nonzero magnetic mass, screening the color magnetic
interaction.  For completeness, we give here an elementary overview of the
approach, in the spirit of non-relativistic quantum mechanics. Readers who
need a more formal, rigorous treatment should consult Ref.\ \cite{EHS}. In
this section we will follow Refs.\ \cite{ARW,RSSV} and consider a theory
of quarks interacting via a local four-fermion interaction.  The aim is to
give an overview of the conventional RG approach before tackling the more
difficult problem of a long-range interaction.  Keeping in mind that
one-gluon exchange conserves helicity, we will for simplicity consider
only left-handed fermions.

Let us imagine a Fermi gas of massless quarks with a chemical potential
$\mu$.  In the ideal gas approximation, all energy levels below the Fermi
surface $|\p|=\mu$ are filled.  The energy will be measured relative to
the Fermi surface, so we introduce $\epsilon_\p=p-\mu$.

The RG procedure works as follows.  At any given step, the effective
theory contains only the fermion degrees of freedom located in a thin
shell surrounding the Fermi surface.  These fermions have
$|\epsilon_\p|<\delta$.  All others have been integrated out.  The only
relevant interaction between the fermions is the scattering of pairs with
opposite momenta \cite{RG,EHS}.  Let us introduce the scattering amplitude
from a pair with momenta $(\p,-\p)$ to another pair with momenta
$(\k,-\k)$,
\[
  f(\theta)\equiv f(\p,\k) = T(\p,-\p\to\k,-\k)
\]
To avoid complications with statistics, we will assume that the two
particles are of different flavors.  Near the Fermi surface, the amplitude
depends only on the angle $\theta$ between $\p$ and $\k$.  A positive $f$
corresponds to a repulsive interaction, and a negative $f$ means
attraction between particles with opposite momenta. 

In the spirit of the Wilson renormalization group, let us now integrate
out all fermion states with $e^{-1}\delta<|\epsilon_\p|<\delta$. According
to quantum mechanics, the scattering through virtual states in this region
gives a correction to the scattering amplitude.  To account for these
virtual processes, we need to correct the scattering amplitude.  Thus $f$
obtains a correction,
\begin{equation}
  f(\p,\k) \to f(\p,\k) - \sum_i 
  {T(\p,-\p\to i) T(i\to\k,-\k)\over E_i-2\epsilon_\p}
  \label{f=sum}
\end{equation}
where the sum is over all intermediate states $i$ belonging to the sector
of the theory that has been integrated out.  The virtual state $i$ may
have a different energy $E_i$ than the initial energy $2\epsilon_\p$.  We
assume that the initial and final particles are almost exactly located at
the Fermi surface, so $\epsilon_\p=\epsilon_\k=0$. What could be the
states $i$?  To answer this question one notices that the scattering
through an intermediate state can be of two types: 

1. The pair $(\p,-\p)$ can scatter to an intermediate pair $(\p',-\p')$,
which then goes to $(\k,-\k)$. In this case, the intermediate state $i$ is
that with two particle excitations with momenta $\pm\p'$.  The Pauli
principle requires that $\p'$ is located above the Fermi surface.  This
state has $E_i$ = $2\epsilon_{\p'}$ and $T(\p,-\p\to i)=f(\p,\p')$,
$T(i\to\k,-\k)=f(\p',\k)$.

2.  Alternatively, first a pair of particles inside the Fermi sea with
momenta $(\p',-\p')$ can scatter to make the final pair $(\k,-\k)$, and
then the initial pair $(\p,-\p)$ scatters to fill the holes vacated by the
pair $(\p',-\p')$ in the Fermi sphere.  In this case, the intermediate
state $i$ consists of six elementary excitations: four particles with
momenta $\pm\p$ and $\pm\k$, and two holes with momenta $\pm\p'$ located
below the Fermi surface, $p'<\mu$.  In this case, $E_i=-2\epsilon_{\p'}$,
$T(\p,-\p\to i)=f(\p',\k)$, $T(i\to\k,-\k)=f(\p,\p')$. 

It is clear now that Eq.\ (\ref{f=sum}) becomes
\begin{equation}
  f(\p,\k) \to f(\p,\k) - 
  \int_{\p'}{f(\p,\p')f(\p',\k)\over 2|\epsilon_{\p'}|}
  \label{f=sum1}
\end{equation}
where the integration is over all $\p'$ satisfying
$e^{-1}\delta<|p-\mu|<\delta$.  The integral over $|\p'|$ can be taken,
and Eq.\ (\ref{f=sum1}) reads,
\[
  f(\p,\k) \to f(\p,\k) - {\mu^2\over2\pi^2} \int\!{d\hat{\p}'\over4\pi}
  {f(\p,\p')f(\p',\k)}
\]
where the integration is over all directions of $\p'$.  Repeating the RG
procedure many times, one finds that $f$ evolves according to the RG
equation,
\begin{equation}
  {d\over dt} f(\p,\k) = -{\mu^2\over2\pi^2}
   \int\!{d\hat{\p}'\over4\pi} {f(\p,\p')f(\p',\k)}
  \label{dftheta/dt}
\end{equation}
where $t=-\ln\delta$ goes to $+\infty$ as one approaches the Fermi
surface.  Eq.\ (\ref{dftheta/dt}) describes the RG evolution of the
scattering amplitude on the Fermi surface.

It is convenient to expand the scattering amplitude over partial waves,
\[
  f(\theta) = \sum_{l=0}^\infty (2l+1)f_lP_l(\cos\theta)
\]
or, inversely,
$f_l={1\over2}\int_0^\pi\!d\theta\,\sin\theta\,P_l(\cos\theta)f(\theta)$. 
Using a well known property of the Legendre polynomials, we find that the
partial-wave amplitudes $f_l$ evolve independently,
\begin{equation}
  {df_l\over dt} = -{\mu^2\over2\pi^2} f_l^2
  \label{dfl/dt}
\end{equation}
The solution to Eq.\ (\ref{dfl/dt}) is
\[
  f_l(t) = {f_l(0)\over1+{\mu^2\over2\pi^2}f_l(0)t}
\]
We see that if at $t=0$ all $f_l>0$, which means that the interaction is
repulsive in all channels, then the four-fermion interaction vanishes at
the Fermi surface.  However, if one of $f_l(0)$ is negative, it will
develop a singularity (the Landau pole) at $t=-{2\pi^2\over\mu^2f(0)}$. 
The Landau pole is reached first by the channel having the largest
negative $f_l(0)$.  This singularity is nothing but the manifestation of
the BCS instability of the Fermi surface with respect to any attractive
interaction.  The BCS gap is proportional to the energy scale at which the
Landau pole is reached,
\[
  \Delta \sim \exp \biggl( -{2\pi^2\over\mu^2f(0)} \biggr)
\]

Let us reproduce some results obtained in Ref.\ \cite{EHS}.  Let us take
the interaction in the form
$G^0(\bar{\psi}\gamma^0\psi)^2+G^i(\bar{\psi}\gamma^i\psi)^2$, where $G^0$
and $G^i$ are two independent constants.  The tree-level scattering
amplitude $\p,-\p\to\k,-\k$ between two left-handed particles arising from
this interaction is
\begin{equation}
  f(\theta) = 2\biggl[G^0\cos^2{\theta\over2}-
              G^i\biggl(\cos^2{\theta\over2}+2\sin^2{\theta\over2}\biggr)
              \biggr]
  \label{fthetaEHS}
\end{equation}
which contains only the $s$-wave and $p$-wave terms, with $f_0=G^0-3G^i$
and $f_1={1\over3}(G^0+G^i)$.  Therefore, $G^0$ and $G^i$ can be said to
run according to the following equations,
\begin{eqnarray*}
  {d\over dt}(G^0-3G^i) & = & -{\mu^2\over2\pi^2} (G^0-3G^i)^2 \\
  {d\over dt}(G^0+G^i) & = & -{\mu^2\over6\pi^2} (G^0+G^i)^2
\end{eqnarray*}
which constitute a subset of the equations found in Ref.\ \cite{EHS}.

As a toy model mimicking the real one-gluon exchange, one can follow Ref.\
\cite{locking} and take the interaction of the form
$-{g^2\over3\Lambda^2}(\bar\psi\gamma_\mu\psi)^2$.  Here $\Lambda$ should
be thought of as the typical momentum of the exchanged gluon, and
${g^2\over3}$ is the effective coupling in the color $\bar{\bf 3}$ channel
\cite{BailinLove,EHS}, where the superconductivity effect is usually
strongest. This corresponds to $G^0=-G^i=-{g^2\over3\Lambda^2}$. The
interaction is most attractive in the $s$-channel and a BCS state is
formed with the gap
\begin{equation}
  \Delta \sim \exp\biggl(-{3\pi^2\Lambda^2\over2\mu^2g^2}\biggr)
  \label{toygap}
\end{equation}
The gap has the $e^{-c/g^2}$ dependence on the coupling $g$.  This
parametric dependence of the gap has been obtained in Ref.\ \cite{EHS} and
is the same as that obtained in variational or mean field treatments of
models with a four-fermion interaction between quarks
\cite{ARW,RSSV,Berges,locking,SchaeferWilczek}.   We will see that in the
true theory, the dependence at asymptotically high densities ($g\to0$)  is
different from that obtained in these toy models.

\section{The problem of the unscreened magnetic field}
\label{sec:mag}

Let us try to naively apply the method developed in Sec.\ \ref{sec:RG} to
high-density QCD.  At the lowest order, the quarks interact by exchanging
one gluon.  The gluon propagator, which is $1/q^2$ in vacuum, is modified
by screening effects.  The static color electric field is subjected to
Debye screening at the distance scale $\md^{-1}$, where $\md\sim g\mu$,
which can be seen by resumming bubble diagrams in the gluon propagator. 
In the magnetic sector, the same resummation yields a magnetic gluon
propagator,
\begin{equation}
  D(q_0,q) = {1\over q^2+{\pi\over2}\md^2{|q_0|\over q}}
  \label{D}
\end{equation}
in the regime $q_0\ll q\ll\mu$.  The term ${\pi\over2}\md^2{|q_0|\over q}$
comes from the Landau damping.  In the static limit $q_0=0$, the magnetic
field is not screened.  If $q_0\neq 0$, the field is said to be
``dynamically screened'' on the scale $q\sim\md^{2/3}q_0^{1/3}$.

Before going further, let us make an important comment on a confusion in
the literature.  This confusion originates from the similarity between the
high-density and high-temperature gauge theories.  The similarity can be
seen in the Debye screening, which occurs at the scale $gT$ at high
temperatures and $g\mu$ at high densities.  At high temperatures, the
magnetic field is not screened at the one-loop level, but develops a
non-perturbative screening of order $g^2T$.  This sometimes leads to an
unjustified assumption that the magnetic field develops a magnetic mass of
order $g^2\mu$ in high-density QCD.

To see why the analogy between high temperatures and high densities breaks
down on the question of the magnetic screening, let us review three
standard ways to interpret the emergence of the magnetic mass in hot gauge
theories.  The first argument is dimensional reduction: the static
large-distance behavior of a gauge theory at high temperatures is the same
as of an effective Euclidean three-dimensional gauge theory.  The latter
is confined on the scale $g^2T$, which means that the magnetic field in
the original theory should also be confined at this scale.  The second
argument is that the high-temperature perturbation theory is IR divergent
for momenta $\alt g^2T$, and only a magnetic mass of this order could make
the perturbation theory finite again. The third way is to notice that, due
to the Bose enhancement, the thermal fluctuations of the gauge field
become so large at the scale $g^2T$ that they are fully non-linear.  The
last argument does not necessarily imply magnetic screening; it just shows
that the existence of the latter at the scale $g^2T$ does not contradict
perturbative results, since at this scale the physics is non-perturbative. 

None of these three arguments can be carried over the case of high
densities.  First, there is no dimensional reduction for gauge theories
with a finite chemical potential.  Second, the perturbation theory for the
long-range magnetic field is infrared finite. Indeed, in vacuum, the
infrared divergences of one-loop graphs come from integrals like
$\int\!d^4q D^2(q)$, where $D(q)\sim q^{-2}$ is the gluon propagator.  At
finite $\mu$, the gluon propagator is modified as in Eq.\ (\ref{D}).  Now
$q_0$ effectively scales like $q^3$ instead of $q$, and by a simple power
counting one sees that the integral is finite in the IR. Therefore
perturbation theory does not break down for momenta of order $g^2\mu$, or
$g^n\mu$ with any $n$, and there is no reason to expect non-perturbative
effects proportional to any finite power of $g$.\footnote{The author
thanks S.Yu.~Khlebnikov for pointing out this argument.  Similar
conclusion has been reached by Pisarski and Rischke \cite{Rischke}.} The
third argument explicitly relies on the large Bose enhancement that takes
place only at finite temperatures; this argument clearly does not work at
finite densities.  Therefore, the magnetic interaction is not screened at
the scale $g^2\mu$ as is occasionally assumed.

Now let us return to the our problem and try to apply the formalism
developed in Sec.\ \ref{sec:RG} to the interaction mediated by one-gluon
exchange.  We will see immediately that we have serious trouble with the
very soft gluons.  Indeed, on the Fermi surface, the tree-level
small-angle ($\theta\ll1$) scattering amplitude, due to one-gluon
exchange, is
\begin{equation}
  f_{\text{tree}}(\theta) =
  - {2g^2\over3}\biggl({1\over\mu^2\theta^2+\md^2}
  + {1\over\mu^2\theta^2}\biggr)
  \label{ftheta}
\end{equation}
The two contributions in the RHS come from the electric and the magnetic
interaction, respectively (again, the factor ${2\over3}$ comes from
considering only the $\bar{\bf 3}$ channel.)  All partial amplitudes
diverge logarithmically.  For example,
\begin{equation}
  f_0 = {1\over2}\int\limits_{q_{\text{min}}/\mu}^\pi\! d\theta\,
  \sin\theta\, P_l(\cos\theta)\, f(\theta)
  \approx -{g^2\over3}\ln {\mu\over q_{\text{min}}}
  \label{fdiverge}
\end{equation}
where $q_{\text{min}}$ is the smallest allowed momentum exchange that one
has to put in by hand to make $f_0$ finite.  Clearly, this IR problem
renders the conventional RG formalism unusable.

Let us also warn against what might seem to be a natural solution to this
IR problem.  If one assumes that a quark condensate is eventually formed,
the magnetic field is screened by the Meissner effect.  One may be tempted
to take the inverse London penetration length $g\Delta$ as the cutoff in
Eq.\ (\ref{fdiverge}), and use the computed value of $f_0$ to find the
gap, thus obtaining a self-consistency condition for $\Delta$.  In this
way one does obtain a gap of order $e^{-c/g}$, where $c$ is some constant. 
However, this approach is flawed, and gives the wrong value of $c$,
because it entirely neglects the screening effect of Landau damping which
turns out to be much stronger than the Meissner effect.  To see this, note
that the condensate smears out the Fermi surface over a scale $\Delta$,
which is the natural energy scale of quasiparticles near the Fermi
surface.  The gluons that such excitations exchange have $q_0\sim\Delta$. 
On these frequencies, the dynamical magnetic screening happens already at
the scale $q\sim\md^{2/3}\Delta^{1/3}$, which is much larger than
$\Delta$.  Therefore, the source of the IR cutoff should be the dynamical
magnetic screening, which already takes place in the normal phase, rather
than the Meissner effect. 

Let us now turn to the central part of this paper, where we will find the
correct RG treatment of the theory with the color magnetic interaction.

\section{Renormalization group for magnetic interactions}
\label{sec:core}

Let us repeat the RG procedure described in Sec.\ \ref{sec:RG}.  At any
given step in the RG evolution, one keeps only fermion modes having
energies smaller than $\delta$.  The start of the RG evolution, $t=0$,
will be taken at $\delta\sim\md$, and the evolution stops when $\delta$ is
of order the gap, so typically $\delta\ll\md$.

At tree level, the fermions interact via one-gluon exchange, characterized
by the momentum of the gluon $(q_0, {\bf q})$.  Since all fermions have
energy less than $\delta$, the energy of the gluon $q_0$ is naturally of
order or less than $\delta$, while the momentum exchange $q$ can be
anywhere between 0 and $2\mu$.

Let us divide the four-fermion interaction that arises from the one-gluon
exchange into ``instantaneous'' and ``non-instantaneous'' parts. The
instantaneous interaction is mediated by the gluons that have momenta
$q\agt q_\delta\equiv\md^{2/3}\delta^{1/3}$.  The Landau damping for these
gluons is negligible, $\md^2{|q_0|\over q}\alt q^2$.  The gluon
propagator, which is now simply $q^{-2}$, does not depend on $q_0$, which
means that the four-fermion interaction they mediate can be considered as
instantaneous.  This part of the interaction is of the familiar type and
will be treated in the conventional way.  In particular, one can
characterize this part by the partial-wave amplitudes $f_l$.  For $q\alt
q_\delta$, the Landau damping can no longer be neglected.  This part of
the interaction has a considerable temporal retardation and should be
treated separately. 

Now let us reduce $\delta$ by a factor of $1/e$ by integrating out fermion
degrees of freedom with energy between $e^{-1}\delta$ and $\delta$. 
During this process the following will occur:

1.  The partial-wave amplitudes $f_l$ obtain the conventional
renormalization, as written in Eq.\ (\ref{dfl/dt}). 

2. One could ask if the non-instantaneous coupling is renormalized during
this integration.  To answer this question, one should compute the
correction to the non-instantaneous interaction that comes from
integrating out the fermion degrees of freedom.  In Appendix
\ref{app:nonrenorm} we explicitly estimate the corresponding one-loop
diagrams and show that the non-in\-stan\-ta\-neous part of the interaction
does not get renormalized. 

3.  Most importantly, and what makes our RG distinctive, {\em part of the
non-instantaneous interaction becomes instantaneous}.  Specifically, the
gluon exchange with $q$ lying in the interval $(e^{-1/3}q_\delta,
q_\delta)$, which was formerly treated as non-instantaneous, now becomes a
part of the instantaneous interaction and contributes to $f_l$.  Simply
speaking, our criterion of what to consider as instantaneous has became
more inclusive, since we are now looking at a smaller energy scale,
corresponding to a larger time scale.

How much of the non-instantaneous part of the interaction transfers to the
instantaneous part during one step of the RG?  According to Eq.\
(\ref{ftheta}) and the non-renormalization of the non-instantaneous
interaction, the increment in $f(\theta)$ has the form,
\begin{equation}
  \Delta f(\theta)=-{2g^2\over3}{1\over\mu^2\theta^2} \qquad
  \text{when $e^{-1/3}{\delta\over\mu}<\theta<{\delta\over\mu}$}
  \label{Deltaf}
\end{equation}
and vanishes outside this window of $\theta$.  For definiteness, let us
concentrate our attention to the $s$-wave amplitude $f_0$.  This amplitude
obtains a constant additive contribution from the soft sector at each RG
step,
\[
  \Delta f_0 = 
  {1\over2}\int\limits_{e^{-1/3}\delta\mu^{-1}}^{\delta\mu^{-1}}\!
  d\theta\,\sin\theta\,\Delta f(\theta) = -{g^2\over9\mu^2}
\]
Therefore, the RG group equation for $f_0$ now becomes,
\begin{equation}
  {d\over dt}f_0 = -{g^2\over9\mu^2} - 
  {\mu^2\over2\pi^2}f_0^2
  \label{RG-eq}
\end{equation}
The second term in the RHS is the familiar term that gives rise to the BCS
effect for short-range interactions.  What is new is the first term, which
takes into account the fact that softer and softer gluon exchanges
contribute to $f_l$.  The non-instantaneous part of the interaction can be
considered as an infinite pool, which continuously replenishes the
instantaneous part during the RG evolution. Clearly, this should speed up
the approach to the Landau pole. 

To secure a solution we also need to specify an initial condition on
$f_0$. Recall that $t=0$ corresponds to $\delta\sim\md$, from Eq.\
(\ref{ftheta}) one finds, to the leading logarithm,
\begin{equation}
  f_0(0) = -{2g^2\over3\mu^2}\ln{1\over g}
  \label{RG-init}
\end{equation}
The solution to Eq.\ (\ref{RG-eq}) with the initial condition
(\ref{RG-init}) is
\[
  f_0(t) = -{\sqrt{2}\pi g\over3\mu^2} \tan \biggl[
  {g\over3\sqrt{2}\pi}\biggl(t+6\ln{1\over g}\biggr)\biggr]
\]
The coupling $f_0$ hits the Landau pole when the argument of the tangent
is equal ${\pi\over2}$.  This happens when
\[
  t={3\pi^2\over\sqrt{2}g} - 6\ln{1\over g}
\]
The Fermi liquid description, thus, breaks down at the energy scale
\begin{equation} 
  \Delta \sim\md e^{-t} \sim \mu g^{-5}\exp\biggl(
  -{3\pi^2\over\sqrt{2}g}\biggr)
  \label{answer}
\end{equation}
which will be interpreted as the scale of the gap.  Notice that the gap is
proportional to $e^{-c/g}$, which is parametrically larger than the naive
estimate $e^{-c/g^2}$ at small $g$.  The reason for this enhancement is
obviously the singularity of the magnetic interaction.

Strictly speaking, the RG calculation does not tell us that $\Delta$ is
the value of the gap.  In fact, the RG merely indicates that the normal
Fermi liquid behavior breaks down at the scale of $\Delta$.  To confirm
that $\Delta$ is the gap, one needs to use some alternative approach.  In
Appendix \ref{app:Etheory}, by making use of the Eliashberg equation,
borrowed from the physics of electron-phonon systems, we verify that the
gap $\Delta$ is in fact proportional to $\exp(-{3\pi^2\over\sqrt{2}g})$.

\section{Remarks and conclusion}
\label{sec:conclusion}

We have found that the ground state of the system of quarks, interacting
via one-gluon exchange, is basically a BCS-type superconductor, and found
the weak-coupling behavior of the gap using the RG approach, appropriate
in the presence of a long-range magnetic interaction.  We found
$\Delta\sim g^{-5}e^{-c/g}$, $c={3\pi^2\over\sqrt{2}}$, not $\Delta\sim
e^{-c/g^2}$ as in conventional BCS theory.  Thus, at least in the $g\to0$
limit, the superconductivity gap is larger than in all previous estimates
\cite{BailinLove,ARW,RSSV,Berges,locking,SchaeferWilczek,EHS}.  Here let
us make a few remarks on our calculation. 

{\em Wave function renormalization and non-Fermi-liquid behaviors}.  A
similar problem of fermions interacting via an unscreened $U(1)$ magnetic
field has also generated considerable interest in condensed matter
physics.  The fermions may be the valence electrons in metals
\cite{Holstein}, in which case the $U(1)$ field is the magnetic component
of real electromagnetism, or some effective degrees of freedom in
low-dimension strongly correlated electron systems \cite{Polchinski},
where the $U(1)$ gauge interaction could be generated as an effective
interaction.  In metals, the magnetic interaction is repulsive for a pair
located on the opposite sides of the Fermi surface, so it does not lead to
BCS superconductivity.  However, interesting phenomena may still arise
from this repulsive interaction.  In particular, in has been shown
\cite{Holstein} that the weak magnetic interaction leads to the breakdown
of the Landau theory of Fermi liquid, typically at extremely low
temperatures. One could ask whether the effects leading to this non-Fermi
liquid behavior would modify our calculations.

The basic idea is that the fermion wave function gets a large
renormalization near the Fermi surface from the magnetic interaction.  To
one-loop level, the renormalization of the wave function is
\cite{Holstein}
\begin{equation}
  Z^{-1}(q_0) = 1 + \text{const}\cdot g^2 \ln {\mu\over q_0}
  \label{Z}
\end{equation}
If one goes arbitrarily close to the Fermi surface, the wave function
renormalization $Z$ tends to 0, which means that the discontinuity of the
distribution function at the Fermi surface disappears, thus signaling a
deviation from Landau's Fermi liquid theory.  However, in our case, the
BCS effect is already essential at $q_0\sim\Delta$.  Recalling that
$\Delta\sim e^{-c/g}$, we find from Eq.\ (\ref{Z}) that $Z$ remains close
to 1, $Z=1+O(g)$.  Therefore, one can safely ignore the renormalization of
the wave function, and there is no chance for non-Fermi-liquid
behavior other than BCS-type superconductivity to manifest itself.

We note here that it was suggested that an attractive magnetic-type
interaction may arise in various situations in condensed matter physics,
for example in double-layer electron systems.  A BCS gap may emerge in
such systems.  Due to a stronger singularity of the magnetic interaction
in 2d, the gap is found to be proportional to a power of the coupling
constant, rather than being exponential \cite{2d}.

{\em The very soft magnetic gluons}.  In our RG approach, even after the
final step of the RG evolution, the interaction still contains a
non-instantaneous sector, which is carried by the magnetic gluons with
energy $q_0\alt\Delta$ and momentum $q\alt\md^{2/3}\Delta^{1/3}$.  The BCS
effect is due to the instantaneous interaction, but one may ask whether
the remaining non-instantaneous interaction could destroy the BCS state. 
Here we give an (admittedly crude) argument as to why this cannot happen. 

The magnetic modes mediating the non-instantaneous interaction have their
intrinsic time scale $q_0^{-1}\agt\Delta^{-1}$.  Therefore, these modes
can be considered as static during the typical time scale of the system,
$\Delta^{-1}$.  The question is now whether this random, static magnetic
field could destroy the superconducting state.  Let us estimate the total
strength of the quantum fluctuations of the magnetic field.  It is roughly
\[
  B^2 \sim \int\limits^{q_\delta}\!d\q\, q^2|A(\q)|^2 \sim 
  \int\!dq_0\,d\q\, q^2{1\over q^2+{\pi\over2}\md^2{|q_0|\over q}} 
  \sim\md^2\Delta^2
\]
so the typical value of the fluctuating magnetic field is $\md\Delta\sim
g\mu\Delta$. This should be compared with the critical magnetic field,
which is of order $\mu\Delta$.  We conclude that at weak coupling, the
almost static quantum fluctuations of the magnetic field are too small to
destroy the superconducting state. 

{\em The possibility of breaking rotational symmetry}. In our treatment of
the RG equation, we have concentrated our attention on the $s$-wave
coupling $f_0$.  However, we could equally consider higher partial waves. 
Let us take arbitrary $l$ and try to rewrite Eq.\ (\ref{RG-eq}) for $f_l$. 
According to Eq.\ (\ref{dfl/dt}), the $f^2$ part in the evolution of $f_l$
has the same coefficient as for $f_0$.  The constant part depends on how
much the non-instantaneous sector throws out to the $l$-channel of the
instantaneous sector at each step of the RG evolution.  To find out, one
should make the partial-wave expansion of the function $\Delta f$ in Eq.\
(\ref{Deltaf}).  Since $\Delta f(\theta)$ is concentrated on values of
$\theta$ near 0, the partial-wave coefficients almost do not depend on
$l$, provided that $l$ is not parametric on $g$.  Indeed,
\[
  \Delta f_l = {1\over2} 
  \int\limits_{e^{-1/3}\delta\mu^{-1}}^{\delta\mu^{-1}}\!
  d\theta\,\sin\theta\, P_l(\cos\theta)\, \Delta f(\theta) \approx
  {g^2\over9\mu^2}P_l(1)={g^2\over9\mu^2}
\]
since the $P_l$ are normalized so that $P_l(1)=1$.  Therefore, the RG
equation for $f_l$ is identical to that for $f_0$.  The initial value for
$f_l$ is also independent of $l$, since it comes from the partial-wave
expansion of the function $f_{\text{tree}}(\theta)$ in Eq.\ (\ref{ftheta}) 
which also peaks near $\theta=0$.  Therefore, to leading order, the RG
does not discriminate between channels with different angular momenta.  If
a condensate with nonzero angular momentum is formed, the rotational
symmetry is broken, like in the A phase of He$^3$. 

However, at any finite $g$, the coincidence of the RG evolution of $f_l$
with different $l$ is not exact.  Moreover, the RG approach does not give
us the value of the gap, or the energy of the ground state, but merely
yields the typical scale of the gap.  Therefore, the two gaps with
different $l$ may have the same asymptotic behavior, but with different
numerical coefficients, and the corresponding superconducting states may
have different energies.  At this stage, the natural assumption seems to
be that the state with $l=0$ is favored, and the ground state does not
break rotational symmetry, but the question of forming a condensate with
nonzero angular momentum of the Cooper pair should be investigated in a
more careful manner.  We defer this question to future work.  We have also
left out the possibility of a numerical estimation of the gap at moderate
densities, which is a very interesting question from the phenomenological
point of view.  Presumably, a reasonable estimation could be found by
solving the Eliashberg equation of the type described in Appendix
\ref{app:Etheory}.  Nor did we try to solve the problem at finite
temperatures \cite{Berges}.

\acknowledgments

The author thanks J.~Berges, S.~Khlebnikov, P.A.~Lee, and K.~Rajagopal for
stimulating discussions at various stages of this work.  I especially
thank K.~Rajagopal for suggesting this problem to me.  This work is
supported in part by funds provided by the U.S.\ Department of Energy
(D.O.E.) under cooperative research agreement \#DF-FC02-94ER40818. 

\appendix

\section{Non-renormalization of the non-instantaneous interaction}
\label{app:nonrenorm}

Here we present an explicit check that the soft sector of the theory is
not renormalized during the RG flow.  Let us consider one-loop corrections
to the scattering amplitude $\p,-\p\to\p,-\p$.  There are two Feynman
diagrams to be considered:
$$
      \def\epsfsize #1#2{1.00#1}
      \epsfbox{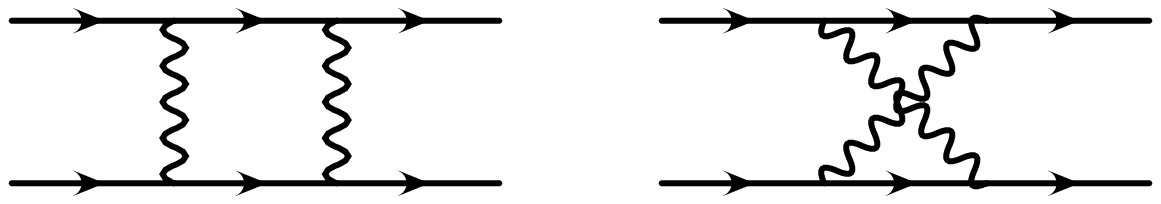}
$$
The contributions from the two graphs are of the same order, let us
evaluate only the first graph, which is of order
\begin{equation}
  g^4\int\!dk_0\,d\k\, {1\over k_0^2+\epsilon_{\p+\k}^2}\cdot
  {1\over(k^2+{\pi\over2}\md^2{|k_0|\over k})^2}
  \label{app:graph}
\end{equation}
Notice that the momentum of the internal fermion lines should be inside
the shell $e^{-1}\delta<|\epsilon_{\p+\q}|<\delta$.  Both gluon lines are
supposed to belong to the non-instantaneous sector, so
$k\alt\md^{2/3}\delta^{1/3}$. The integral over $k_0$ is dominated by
$k_0\alt\delta$. Therefore, the integral in Eq.\ (\ref{app:graph}) is of
order
\[
  g^4 \cdot \delta \cdot q_\delta^2\delta \cdot {1\over\delta^2}\cdot
  {1\over q_\delta^4} \sim {g^4\over q_\delta^2}
\]
Therefore, the correction is of order $g^4q_\delta^{-2}$.

Now assume that the external momenta of the final particles are slightly
different from those of the initial particles, and the difference is $q$.
The tree-level amplitude is of order $g^2q^{-2}$.  During each step of the
RG evolution, this amplitude receives a correction of order $g^4
q_\delta^{-2}$, where $\delta$ is the moving RG scale.  During the part of
the RG evolution when the interaction is non-instantaneous, $q_\delta\agt
q$, and one sees that all the accumulated correction ($\sim g^4q^{-2}$) is
still smaller than the tree amplitude by a factor of $g^2$.  One concludes
therefore that there is no substantial renormalization of the
non-instantaneous interaction during the RG evolution.

\section{The Eliashberg theory}
\label{app:Etheory}

The Eliashberg equations \cite{Mahan} can be considered as the
generalization of the BCS gap equation to the case of a non-local
interaction.  Here we will be trying to reproduce only the leading
exponential behavior $\exp(-{3\pi^2\over\sqrt{2}g})$ of the gap, but not
the power ($g^{-5}$) part.  Presumably, a more careful treatment of the
Eliashberg equation should reproduce the subleading $g^{-5}$ factor and
give an estimate for the numerical coefficient.

The generalization of the gap equation of the type written in Ref.\
\cite{locking} to the case of the non-local magnetic interaction
is\footnote{The coefficient ${2\over3}$ in Eq.\ (\ref{Eliashberg}) can be
understood as follows.  If one replaces the gluon propagator in Eq.\
(\ref{Eliashberg}) by ${1\over\Lambda^2}$, Eq.\ (\ref{Eliashberg}) gives
the same gap as Eq.\ (\ref{toygap}) if the numerical coefficient in the
RHS is ${4\over3}$.  The coefficient one should put in the RHS of Eq.\
(\ref{Eliashberg}) should be twice smaller than that, due to the screening
of the electric field, which reduces the effective coupling by a factor of
2.  For the color-flavor locking scheme \cite{locking}, there should be
two equation for two gaps; in weak coupling they can be written as one
equation (\ref{Eliashberg}) which is valid to the leading order.}
\begin{equation}
  \Delta(p_0) = {2\over3}g^2\int\!{d^4q\over(2\pi)^4}
  {1\over|\p-\q|^2+\mu^2{|p_0-q_0|\over|\p-\q|}}\cdot
  {\Delta(q_0)\over q_0^2+\epsilon_\q^2+\Delta^2(q_0)}
  \label{Eliashberg}
\end{equation}

Notice that, as in the usual Eliashberg theory, the gap is a function of
$p_0$ only but not of $\p$.  Indeed, the dependence of the RHS of Eq.\
(\ref{Eliashberg}) on $\p$ is only in the gluon propagator, and if $\p$ is
near the Fermi surface any change of $\p$ can be compensated by a rotation
of $\q$.  The conventional Eliashberg equations are actually a set of
equations for $\Delta(p_0)$ and $Z(p_0)$, where $Z(p_0)$ is the wave
function renormalization, but as explained in Sec.\ \ref{sec:conclusion},
$Z(p_0)\approx1$. Notice also that in Eq.\ (\ref{Eliashberg}) we write
$\mu^2$ instead of ${\pi\over2}\md^2$, the reason is that we will be
working with exponential scales so the difference between these two
coefficients will not affect the leading exponent.  Integrating over $\q$,
one finds,
\begin{equation}
  \Delta(p_0) = {g^2\over18\pi^2} \int\limits_0^\infty dq_0\,
  \ln{\mu\over|p_0-q_0|}\cdot {\Delta(q_0)\over\sqrt{q_0^2+\Delta^2(q_0)}}
  \label{Eliash2}
\end{equation}
where only the leading logarithm is written.  To the leading log, the
integral in Eq.\ (\ref{Eliash2}) can be split into two regions,
$0<q_0<p_0$ and $p_0<q_0<\mu$, where in the first
$\ln{\mu\over|p_0-q_0|}\approx\ln{\mu\over p_0}$ and in the second
$\ln{\mu\over|p_0-q_0|}\approx\ln{\mu\over q_0}$.  Introducing the
logarithmic scales
\[
  x=\ln{\mu\over p_0},\quad y=\ln{\mu\over q_0},\quad
  x_0=\ln{\mu\over\Delta_0}
\]
where $\Delta_0=\Delta(p_0=0)$, the Eliashberg equation becomes,
\begin{equation}
  \Delta(x)={g^2\over18\pi^2} \biggl( x\int\limits_x^{x_0}\! dy\,\Delta(y)
  + \int\limits_0^x\! dy\,y\Delta(y) \biggr)
  \label{EliashD}
\end{equation}
Differentiating Eq.\ (\ref{EliashD}) with respect to $x$ twice, one finds,
\begin{equation}
  \Delta''(x) = -{g^2\over18\pi^2}\Delta(x)
  \label{EliashD2}
\end{equation}
As the boundary conditions, from Eq.\ (\ref{EliashD}) one can check that
$\Delta(0)=0$ and $\Delta'(x_0)=0$.  The solution to Eq.\ (\ref{EliashD2})
is,
\[
  \Delta(x) = \Delta_0\sin\biggl( {g\over3\sqrt{2}\pi} x\biggr)
\]
To satisfy the boundary condition at $x_0$, one requires
${g\over3\sqrt{2}\pi} x_0={\pi\over2}$, from which one finds
$x_0={3\pi^2\over\sqrt{2}g}$, and the gap at small energies is
$\Delta_0\sim e^{-x_0}\sim \exp(-{3\pi^2\over\sqrt{2}g})$, reproducing
the leading exponential behavior of our RG result.  This is the minimal
energy cost to create a fermion excitation.  The energy-dependent gap
$\Delta(p_0)$ is
\[
  \Delta(p_0) = \Delta_0\sin\biggl({g\over3\sqrt{2}\pi}
  \ln{\mu\over p_0}\biggr)
\]
for $p_0\agt\Delta_0\sim\exp(-{3\pi^2\over\sqrt{2}g})$.


\begin{references}

\bibitem{cp} J.C.~Collins and M.J.~Perry, Phys. Rev. Lett. {\bf 34}, 1353
(1975).

\bibitem{gwp} D.J.~Gross and F.~Wilczek, Phys. Rev. Lett. {\bf 30}, 1343
(1973); H.~Politzer, Phys. Rev. Lett. {\bf 30}, 1346 (1973).

\bibitem{BailinLove} D.~Bailin and A.~Love, Phys. Rept. {\bf 107}, 325
(1984), and references there in.

\bibitem{ARW} M.~Alford, K.~Rajagopal, and F.~Wilczek, Phys. Lett. B {\bf
422}, 247 (1998).

\bibitem{RSSV} R.~Rapp, T.~Sch\"afer, E.V.~Shuryak, and M.~Velkovsky,
Phys. Rev. Lett. {\bf 81}, 53 (1998).

\bibitem{Berges} J.~Berges and K.~Rajagopal, {\tt hep-ph/9804233}.

\bibitem{locking} M.~Alford, K.~Rajagopal, and F.~Wilczek, {\tt
hep-ph/9804403}.

\bibitem{SchaeferWilczek} T.~Sch\"afer and F.~Wilczek, {\tt
hep-ph/9811473}.

\bibitem{RG} G.~Benfatto and G.~Galavotti, {\em Renormalization Group},
Princeton University Press, Princeton, 1995; R.~Shankar, Rev. Mod. Phys.
{\bf 66}, 129 (1993); J.~Polchinski, {\tt hep-th/9210046}.

\bibitem{EHS} N.~Evans, S.D.H.~Hsu, and M.~Schwetz, {\tt hep-ph/9808444},
{\tt hep-ph/9810514}; T.~Sch\"afer and F.~Wilczek, {\tt hep-ph/9810509}.

\bibitem{Rischke} R.~Pisarski and D.H.~Rischke, {\tt nucl-th/9811104}.

\bibitem{Holstein} T.~Holstein, R.E.~Norton, and P.~Pincus, Phys. Rev. B
{\bf 6}, 2649 (1973); M.Yu.~Reizer, Phys. Rev. B {\bf 40}, 11571 (1988);
M.Yu.~Reizer, Phys. Rev. B {\bf 44}, 5476 (1991).

\bibitem{Polchinski} See, e.g., C.~Nayak and F.~Wilczek, Nucl. Phys. B
{\bf 417}, 359 (1994); J.~Polchinski, Nucl. Phys. B {\bf 422}, 617 (1994);
B.L.~Altshuler, L.B.~Ioffe, and A.J.~Millis, Phys. Rev. B {\bf 50}, 14048
(1994); 

\bibitem{2d} B.~Normand, P.A.~Lee, and N.~Nagaosa, Physica (Amsterdam)
{\bf 185-189C}, 1479 (1991); D.V.~Khveshchenko, Phys. Rev. B {\bf 47},
3446 (1993); N.E.~Bonesteel, I.A.~McDonald, and C.~Nayak, Phys. Rev. Lett.
{\bf 77}, 3009 (1996).

\bibitem{Mahan} See, e.g., G.~Mahan, {\em Many-Particle Physics}, 2nd Ed.,
Plenum, New York, 1990, Section 9.7. 


\end{references}
\end{document}